\documentclass{elsart}

\usepackage{natbib}

\usepackage{graphicx}

\usepackage{amssymb}

\begin{document}

\begin{frontmatter}



\title{Recent Results from the MAXIMA Experiment}


\author[1]{Andrew H. Jaffe}
\author[2]{Matthew Abroe}
\author[3,4]{Julian Borrill}
\author[5]{Jeff Collins}
\author[6]{Pedro Ferreira}
\author[2]{Shaul Hanany}
\author[2]{Brad Johnson}
\author[5]{Adrian T. Lee}
\author[2]{Tomotake Matsumura}
\author[5]{Bahman Rabii}
\author[2]{Tom Renbarger}
\author[5]{Paul Richards}
\author[4,5]{George F. Smoot}
\author[3,4,5]{Radek Stompor}
\author[5]{Huan Tran}
\author[5]{Celeste Winant}
\author[7]{Jiun-Huei Proty Wu}

\address[1]{Imperial College London}
\address[2]{University of Minnesota}
\address[3]{National Energy Resource Supercomputing Center}
\address[4]{Lawrence Berkeley National Laboratory}
\address[5]{University of California, Berkeley}
\address[6]{University of Oxford}
\address[7]{National Taiwan University}

\begin{abstract}
  MAXIMA is a balloon-borne platform for measuring the anisotropy of
  the Cosmic Microwave Background (CMB). It has measured the CMB power
  spectrum with a ten-arcminute FWHM beam, corresponding to a detection
  of the power spectrum out to spherical harmonic multipole
  $\ell\sim1000$. The spectrum is consistent with a flat Universe with a
  nearly scale-invariant initial spectrum of adiabatic density
  fluctuations. Moreover, the MAXIMA data are free from any notable
  non-Gaussian contamination and from foreground dust emission. In the
  same region, the WMAP experiment observes the same structure as that
  observed by MAXIMA, as evinced by analysis of both maps and power
  spectra. The next step in the evolution of the MAXIMA program is
  MAXIPOL, which will observe the polarization of the CMB with
  comparable resolution and high sensitivity over a small patch of the
  sky.
\end{abstract}




\end{frontmatter}

\section{Introduction}
\label{sec:intro}

In 2000, MAXIMA-I \citep{Hanany00} [along with BOOMERANG-98
\citep{debern00}, with which it shares considerable personnel and
technology] was one of the first CMB experiments to detect anisotropy
over a range of angular scales from several degrees down to a resolution
of ten arcminutes. The pattern of temperature anisotropy, with a
characteristic scale of roughly one degree, was as expected from a flat
Universe with a nearly scale-invariant initial power spectrum of
adiabatic perturbations, as predicted by, for example, Inflation. This
was confirmed in a combined analysis \citep{maxiboom01} and by
comparison with the many other experiments that have followed,
including the recent results from WMAP (see other contributions to
this volume).

In this Proceedings, we will give an overview of the MAXIMA experimental
program, report on results to date from MAXIMA, including cosmological
results from the analysis of primary anisotropy as well as efforts to
detect departures from a Gaussian distribution and contributions from
foreground emission. We will discuss comparisons with results from the
WMAP experiment, and ongoing plans to use MAXIMA as a CMB polarimeter:
MAXIPOL.

The MAXIMA program has so far resulted in ten letters and journal papers
published by the team, several works in progress, many conference
proceedings, and three Ph.D. theses (with more coming), in addition to the
many works by other authors using the results. These publications
contain far more information than can be reproduced here.

\section{MAXIMA: Hardware and Strategy}

MAXIMA is a balloon-borne platform carrying a focal plane of spider-web
bolometers cooled to 100 mK. It observes the CMB at 150, 240 and 410 GHz with a
resolution of better than ten arcminutes at all frequencies. The telescope
is an off-axis Gregorian mount with a 1.3 meter primary mirror. The
bolometers have a Noise-Equivalent Temperature (NET) of
80--90$\mu{\mbox{K}}\sqrt{\mbox{sec}}$ at 150 GHz. 

MAXIMA flew on August 2, 1998.  The scan strategy was designed to
revisit the same area of sky over a wide variety of timescales to allow
for a robust investigation of possible systematic errors as well as
reduction of the effects of ``1/f noise''. The primary mirror chops at
0.45 Hz over about $\pm2^\circ$ in azimuth at fixed elevation. The gondola
simultaneously scans in azimuth at 0.02 Hz over an angular scale
designed to result in a roughly $10^\circ\times10^\circ$ patch of sky.
The elevation of the detector was shifted in mid-flight to cover the
same area. During the same flight, observations of Jupiter and the CMB
dipole were carried out for calibration and beam mapping (Jupiter
only). The approximately three hours spent observing a field of
approximately 120 square degrees resulted in a noise per beam of tens
of $\mu$K.

\section{Primary CMB Anisotropy: Maps, Power Spectra and Cosmological Parameters}
\label{sec:primary}

The primary scientific results from MAXIMA are in the form of CMB maps
and power spectra \citep{Hanany00,Lee01}. They were produced by the
MADCAP \citep{MADCAP} program implementing a Bayesian/maximum likelihood
CMB data analysis algorithm \citep{BJK98,BJK00}.

\begin{figure}[htbp]
  \centering
  \includegraphics[width=1\columnwidth]{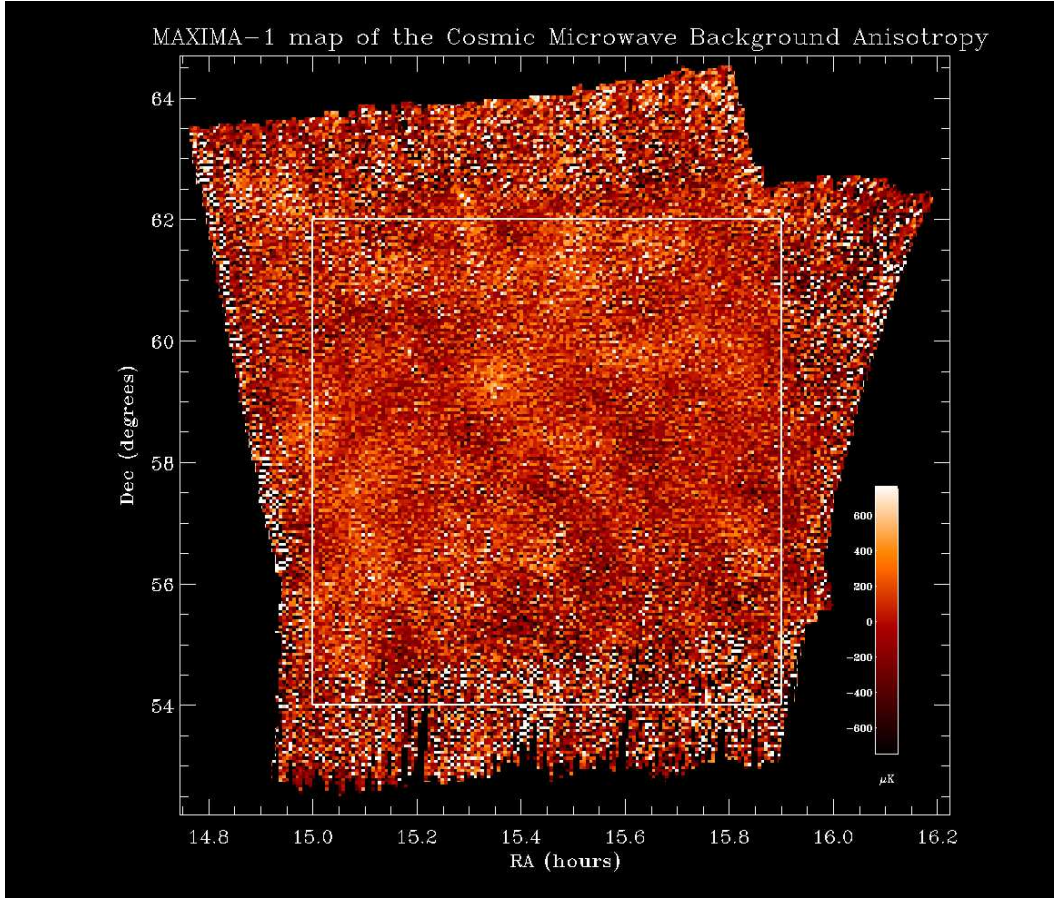}
  \caption{Map of the MAXIMA-1 field. The temperature in $\mu$K is given
  by the colorbar to the right. The central square was used for
  subsequent analysis.}
  \label{fig:map}
\end{figure}

\begin{figure}[htbp]
  \centering
  \includegraphics[width=1\columnwidth]{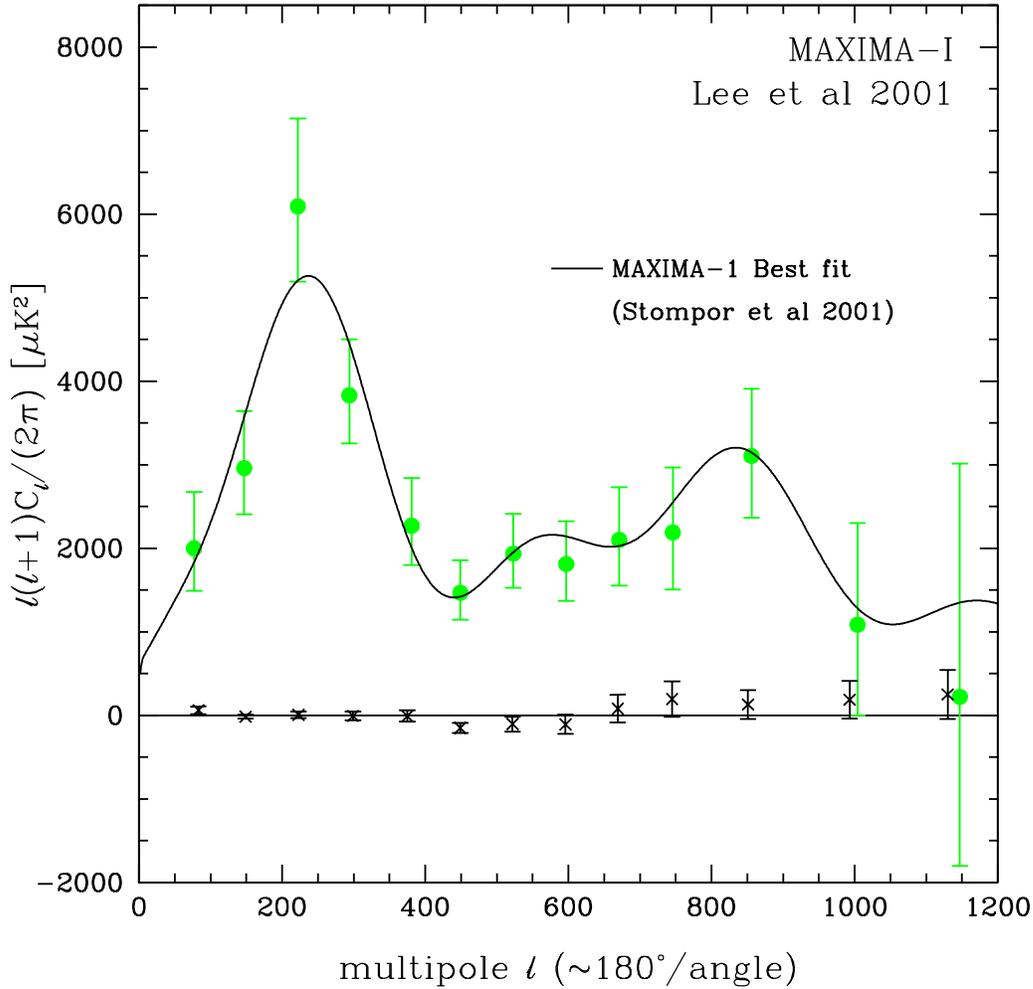}
  \caption{CMB Power spectrum from MAXIMA-1, from \citet{Lee01}. The
    filled circles give the maximum-likelihood solution and errorbars
    (decorrelated using the technique of \citet{BJK98}); the crosses and
  errorbars give an estimate of the power spectrum from two independent
  maps given by different combinations of MAXIMA detectors. The curve is
  a best-fit model from \citet{Stompor01}.}
  \label{fig:Cl}
\end{figure}

This power spectrum was then used to estimate the cosmological
parameters within the paradigm of big-bang models with adiabatic initial
conditions described by a nearly power-law spectrum of primordial
perturbations (i.e., so-called inflationary models). The data prefer a
flat Universe ($\Omega_{\rm tot}\simeq0.9\pm0.2$ at 95\% confidence)
with a nearly scale-invariant spectrum of initial density fluctuations
($n_s\simeq1.08^{+0.22}_{-0.12}$) and a baryon density as determined
from big-bang nucleosynthesis ($\Omega_b h^2\simeq0.33\pm0.13$). These
are all consistent with expectations from the inflationary paradigm and
with determinations of the parameters from other ground- and
balloon-based experiments
\citep[e.g.,][]{DASI2,Pearson02,GraingeVSA2002} and from the WMAP
satellite \citep{WMAP03Spergel}.

In addition to the work by the MAXIMA team itself, these data have also
been used to test the details of quintessence or dark energy
\citep{BalbiQuint01} and Big-Bang nucleosynthesis
\citep{2001PhRvD..63d3004E,2002APh....17...87C}, as well as more exotic
models with massive neutrinos, nontrivial topology, extra dimensions,
isocurvature fluctuations, topological defects, etc. There remains no
compelling evidence from these data for anything beyond the ``standard
cosmology'' discussed above.

\section{Foregrounds \& Non-Gaussianity}

Thus far, no departure from Gaussianity have been detected at a level
that is signficant or would affect these results.  Indeed, in the main
analysis described in the previous section, we assume that the data are
as simple as possible: uncontaminated by foreground emission and with
both signal and noise well-described by Gaussian statistics. These
assumptions can be tested and any departures from them can be
quantified.

The question of non-Gaussianity is discussed elsewhere in these
Proceedings (Heavens 2003) as well as in greater detail in
\citet{Wu01NG}, which considers 82 (not independent) pixel-based tests,
and in \citet{Santos02Bispec,Santos03Bispec}, which calculate the
cosmological bispectrum (3-point function in harmonic space). Thus far,
no significant departures from Gaussianity have been detected.

The MAXIMA field was chosen to be well off of the Galactic plane and in
a region of particularly low dust contrast. Nonetheless, in
\citet{JaffeDust03}, we show that there is a small amount of emission
correlated with the pattern of dust morphology. The amplitude of the
correlated emission is barely detectable, but as expected from
extrapolations of the dust spectrum down to the 150-400 GHz observed by
MAXIMA. This emission has no significant effect on the power spectrum or
cosmological parameters.

\section{MAXIMA and WMAP}

\begin{figure}[htbp]
  \centering \includegraphics[width=1\columnwidth]{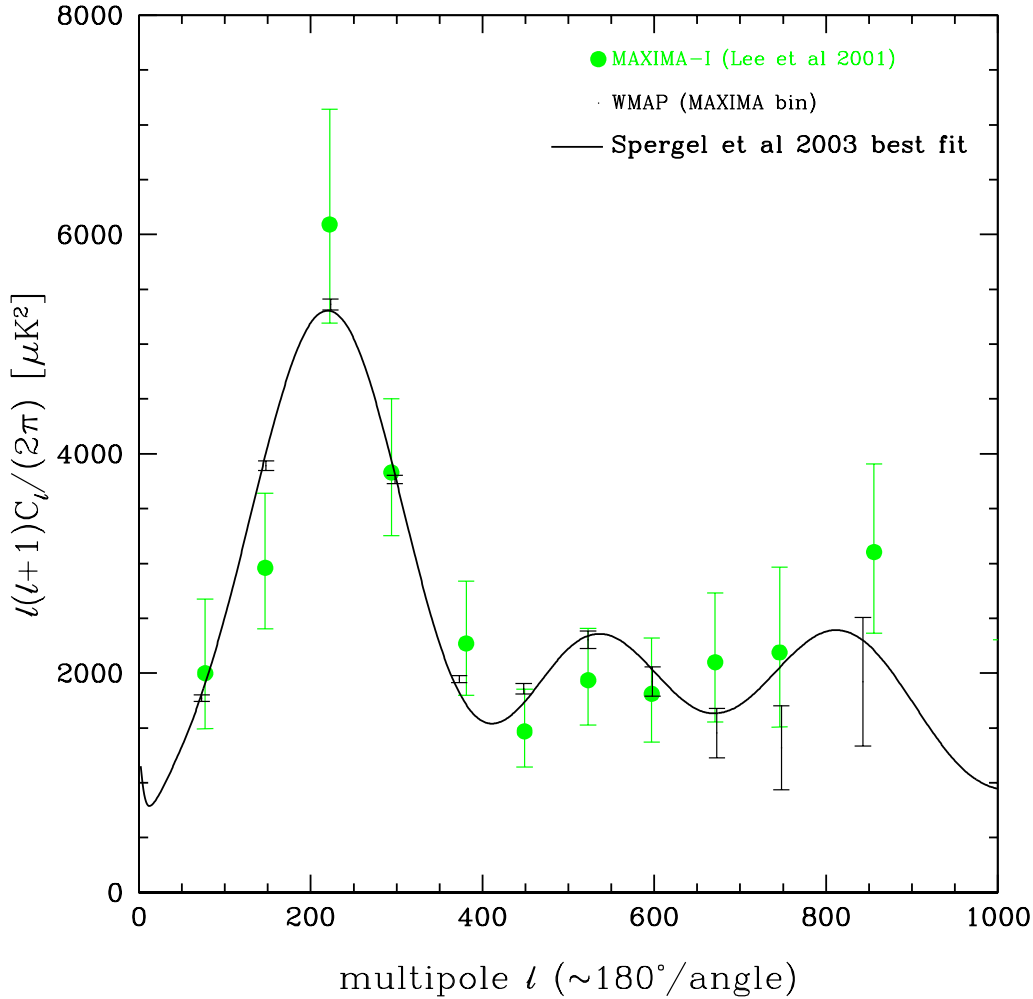}
  \caption{MAXIMA-1 and WMAP power spectra. The filled green circles are
    the MAXIMA data from \cite{Lee01}; the error bars without symbols
    are the WMAP data, rebinned to the same $\ell$s as MAXIMA, weighted
    by the inverse variance. The solid curve is the best fit from the
    WMAP team.}
  \label{fig:WMAP-MAXIMA-Cl}
\end{figure}

\begin{figure}[htbp]
  \centering \includegraphics[width=0.45\columnwidth]{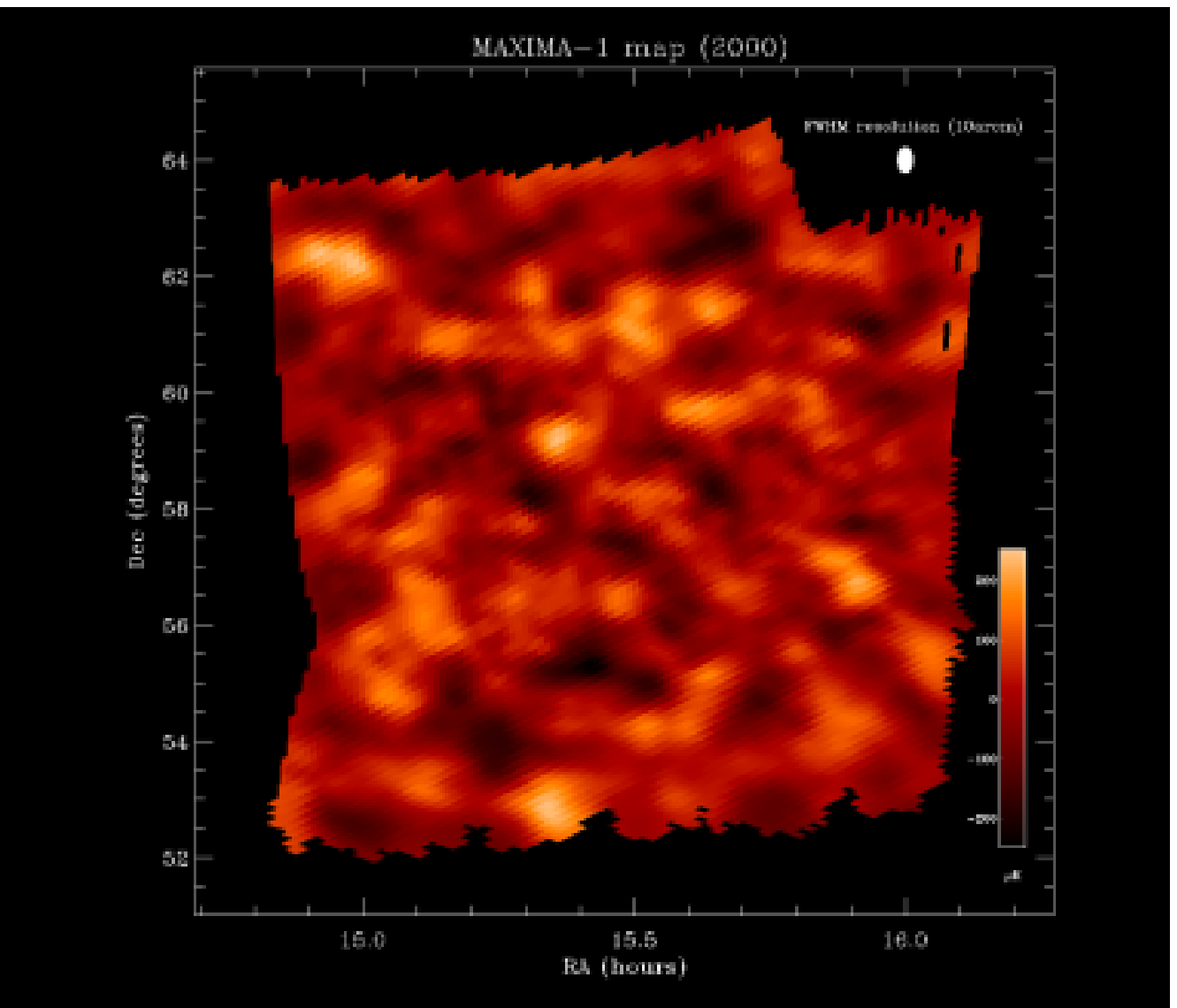}\includegraphics[width=0.45\columnwidth]{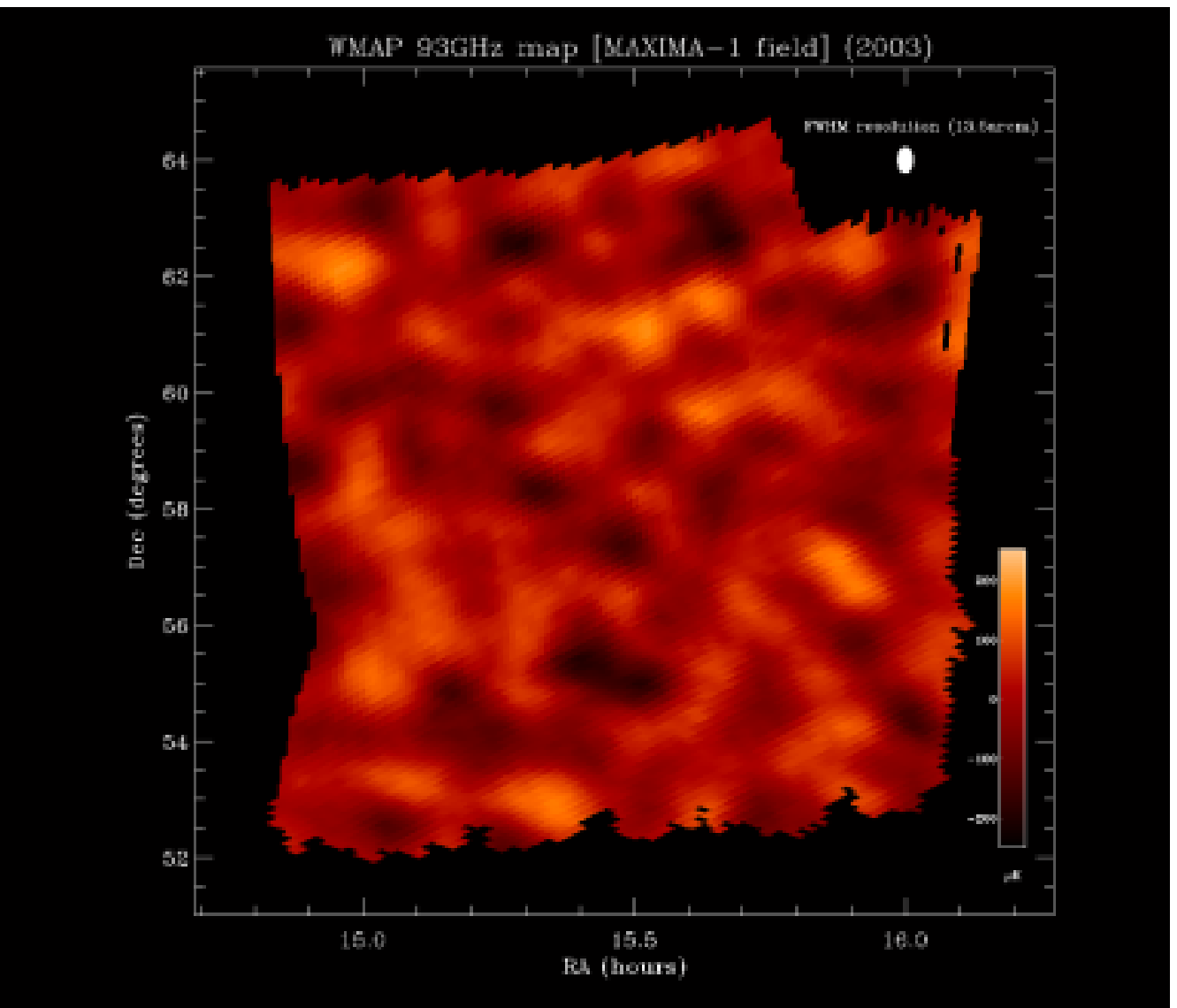}
  \caption{MAXIMA (left) and WMAP (right) maps of the MAXIMA-1
    region. Both maps have been Wiener filtered and smoothed for
    presentations. Note that the MAXIMA map has a resolution of about 10
    arcminutes, the WMAP map about 13 arcminutes.}
  \label{fig:WMAP-MAXIMA-maps}
\end{figure}

In February, 2003, data from the WMAP satellite became available,
providing a benchmark against which the last few years' ground- and
balloon-based CMB results can be tested, down to WMAP's resolution of 13
arcminutes. We will provide a full comparison of the MAXIMA data with
that from the WMAP satellite in Abroe et al (forthcoming). Here, we
concentrate on the comparison of publicly-available data from both
teams. In Figure~\ref{fig:WMAP-MAXIMA-Cl} we compare power spectra.  As
is evident from this figure, the field probed by the MAXIMA data shows
structure with the same statistical properties as, and with higher
resolution than, the all-sky WMAP data. This comparison is borne out by
a side by side visual comparison of the maps of the MAXIMA region
(Figure~\ref{fig:WMAP-MAXIMA-maps}), and with a more detailed comparison
of the overlapping portion of the two datasets as well as MAXIMA data
from other campaigns (Abroe et al, forthcoming).

\section{MAXIPOL}

\begin{figure}[htbp]
  \centering
    \centering \includegraphics[width=1\columnwidth]{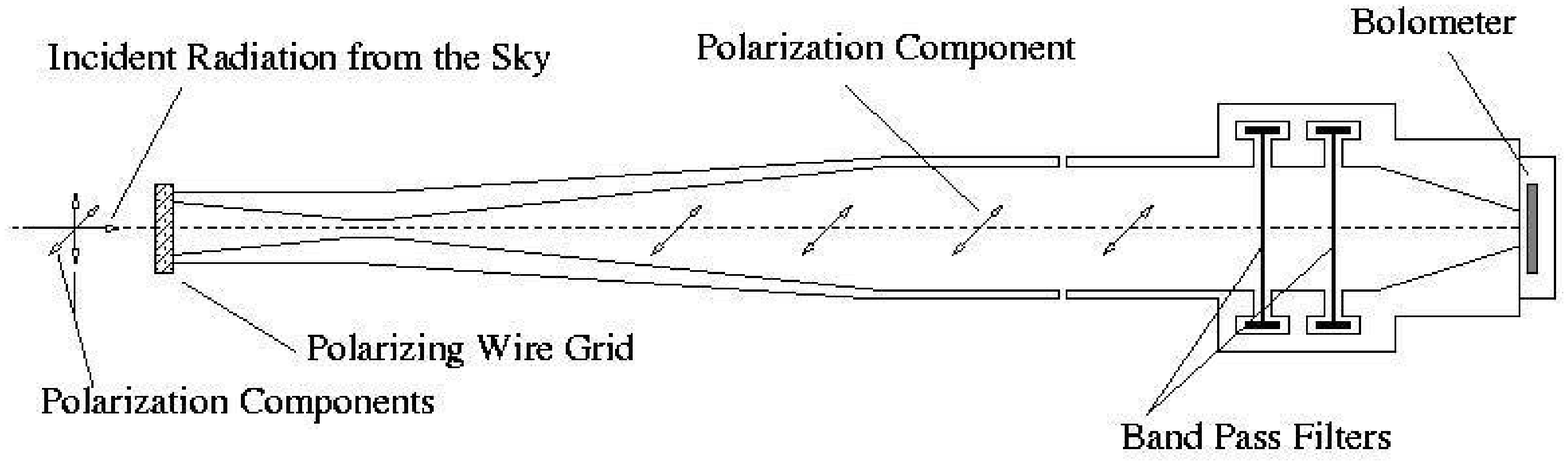}

  \caption{Diagram of the MAXIPOL polarizer, feed horn and bolometer.}
  \label{fig:maxipolhorn}
\end{figure}

The next step in the evolution of the MAXIMA program is MAXIPOL. It
consists of the MAXIMA hardware with additions to make it sensitive to
the polarization of the CMB. A wire grid is placed in front of the
detector horns, and a half-wave plate is rotated at 2~Hz in front (see
Figure~\ref{fig:maxipolhorn}). This means that a single detector can
observe polarization states rotated around a full $180^\circ$ on
timescales much shorter than the drift in detector gain, and adding
another timescale over which we can monitor for systematic effects.

MAXIPOL should have a polarization sensitivity (in $C_\ell$) comparable
to that of WMAP and other recent experiments. It achieves this by
combining the high-resolution MAXIMA optics and the cooled MAXIMA
bolometers with a very deep ``exposure'' of a small area of sky (a few
square degrees).

The MAXIPOL team is happy to report that on 24 May, 2003, just prior to
the submission of these proceedings, the MAXIPOL instrument was
successfully launched for a twenty-five hour flight from Fort Sumner,
New Mexico.

\section{Conclusions}

The MAXIMA program has measured the temperature anisotropy of the Cosmic
Microwave Background over angular scales from ten degrees to ten
arcminutes, and over a factor of several in frequency. It has used these
measurements to determine the CMB power spectrum and the parameters of
the cosmological. It will soon be able to detect the polarization of the
CMB, which will allow a strong check on the underlying cosmological
paradigm as well as the determination of the parameters to higher
precision. Similar technology will be used in the forthcoming Planck
satellite mission as well as future ground- and balloon-based detectors.



\def\nat{Nature}
\def\apjl{Astrophys.\ J.\ Lett.}
\def\apj{Astrophys.\ J.}
\def\prd{Phys.\ Rev.\ D}
\def\mnras{Mon.\ Not.\ R.\ Astr.\ Soc.}



\end{document}